\documentclass[conference]{IEEEtran}
\IEEEoverridecommandlockouts

\usepackage[utf8]{inputenc}
\usepackage{color}
\usepackage{amsmath}
\usepackage{graphicx}
\usepackage[acronym]{glossaries}
  \setacronymstyle{long-short}
  \newacronym{st}{S/T}{Schmitt-Trigger}
\usepackage[siunitx]{circuitikz}
\usepackage{tikz}
  \usetikzlibrary{intersections,external}
  \tikzset{
tablenum/.style={draw,circle,inner sep=0.3mm,thick,font=\bfseries},
cell/.style={draw,minimum width=6mm, minimum height=4mm},
downfast/.pic={\draw [->] (0,0.9mm) -- (0,-1mm);},
upfast/.pic={\draw [<-] (0,1mm) -- (0,-0.9mm);},
miniglitch/.pic={\draw (-0.7mm,-0.9mm) -- ++(0.5mm,0) ..
  controls +(0.5mm,0.5mm) .. ++(1mm,0) -- ++(0.5mm,0);},
glitch/.pic={\draw [rounded corners=0.3mm] (-0.7mm,-0.9mm) -| ++(0.5mm,1.9mm) -|
  ++(0.8mm,-1.9mm) -- ++(0.5mm,0);},
late/.pic={\draw [->,rounded corners=0.5mm] (-0.7mm,0.9mm) -| ++(1.1mm,-1.9mm);},
uplate/.pic={\draw [->,rounded corners=0.5mm] (-0.7mm,-0.9mm) -| ++(1.1mm,1.9mm);},
double/.pic={\draw [rounded corners=0.4mm] (-0.7mm,-0.9mm) -| ++(1.1mm,0.8mm) -| ++(0.8mm,-0.8mm) -- ++(0.3mm,0);
  \draw (-0.7mm,0.5mm) -- ++(2.2mm,0);},
nopic/.pic={},
metaup/.pic={\draw [->,rounded corners=0.7mm] (-0.7mm,0) -- ++(1.1mm,0) -- ++(60:1mm);},
metadown/.pic={\draw [->,rounded corners=0.7mm] (-0.7mm,0) -- ++(1.1mm,0) -- ++(-60:1mm);},
meta/.pic={\draw [->,rounded corners=0.7mm] (-0.7mm,0) -- ++(1.1mm,0) -- ++(0:1mm);},
doubleup/.pic={\draw [->,rounded corners=0.6mm] (-0.7mm,0.2mm) -- ++(0.8mm,0) -- ++(60:0.7mm);
  \draw [->,rounded corners=0.3mm] (-0.7mm,-0.4mm) -- ++(0.8mm,0) -- ++(-80:0.4mm) --
  ++(0.8mm,0) -- ++(60:1.3mm);},
doubledown/.pic={\draw [->,rounded corners=0.6mm] (-0.7mm,-0.2mm) -- ++(0.8mm,0) -- ++(-60:0.7mm);
  \draw [->,rounded corners=0.3mm] (-0.7mm,0.4mm) -- ++(0.8mm,0) -- ++(80:0.4mm) --
  ++(0.8mm,0) -- ++(-60:1.3mm);},
}
\usepackage{pgfplots}
  \usepgfplotslibrary{groupplots}
  \pgfplotsset{
    compat=1.3,
    every axis/.append style={font=\scriptsize\itshape},
    every axis plot/.append style={thick},
    every axis grid/.append style={dotted},
    every tick/.append style={major tick length=1mm},
    height=6.7cm,
  }

\usepackage{textpos}
\usepackage{hyperref}

\newcommand{\figPath}[1]{./#1.pdf}

\newcommand{\tabNum}[1]{\scalebox{0.8}{\tikz[baseline=-1ex]{\node [tablenum] {#1};}}}

\begin{document}

\title{Does Cascading Schmitt-Trigger Stages\\Improve the Metastable Behavior?}
\author{
\IEEEauthorblockN{ Andreas Steininger
\begin{minipage}[c]{1em}
\href{https://orcid.org/0000-0002-3847-1647}{\includegraphics[width=1em]{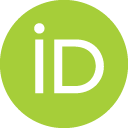}}
\end{minipage}\,
and\: Robert Najvirt\; and\: J\"urgen Maier
\begin{minipage}[c]{1em}
\href{https://orcid.org/0000-0002-0965-5746}{\includegraphics[width=1em]{orcID.png}}
\end{minipage}
}
\IEEEauthorblockA{Vienna University of Technology, 1040 Vienna, Austria\\
\{steininger, rnajvirt, jmaier\}@ecs.tuwien.ac.at}
\thanks{This research was partially supported by the SIC project (grant
  \mbox{P26436-N30}) of the Austrian Science Fund (FWF).}
}

\maketitle

\begin{textblock*}{\textwidth}(0mm,189mm)%
  \footnotesize%
  \textcopyright\ 2016 IEEE.  Personal use of this material is permitted.  Permission from IEEE
  must be obtained for all other uses, in any current or future media, including
  reprinting/republishing this material for advertising or promotional purposes,
  creating new collective works, for resale or redistribution to servers or lists,
  or reuse of any copyrighted component of this work in other works.
\end{textblock*}%
\vspace*{-1em}

\begin{abstract}
Schmitt-Trigger stages are the method of choice for robust discretization of
input voltages with excessive transition times or significant noise.
However, they may suffer from metastability. 
Based on the experience that the cascading of flip-flop stages yields a dramatic 
improvement of their overall metastability hardness, in this paper we elaborate
on the question whether the cascading of Schmitt-Trigger stages can obtain a
similar gain. 

We perform a theoretic analysis that is backed up by an existing metastability
model for a single Schmitt-Trigger stage and elaborate some claims about the
behavior of a Schmitt-Trigger cascade. These claims suggest that the occurrence
of metastability is indeed reduced from the first stage to the second which
suggests an improvement. On the downside, however, it becomes clear that
metastability can still not be completely ruled out, and in some cases the
behavior of the cascade may be less beneficial for a given application, e.g. by
introducing seemingly acausal transitions. We validate our findings by extensive
HSPICE simulations in which we directly cover our most important claims. 
\end{abstract}

\section{Introduction}

In VLSI design two types of discretization are performed to make real-world
quantities ``computable'' in a digital domain. The \textit{values} of originally
analog quantities are transformed from their continuous space into a suitable
digital (i.e. discrete) representation by analog-to-digital converters. 
In the \textit{time domain} a periodic clock signal performs a transformation from
a continuous to a discrete time by sampling these digital logic states at
suitable instances, typically the clock edge. It has been formally proven that
every such transformation from a continuous space to a discrete one inevitably
introduces the risk of metastability~\cite{Mar81}, i.e.\ a prolonged state of
indecision between two discrete values. Consequently one either 
has to allow an unbounded decision time, or expect to occasionally
experience undefined, also known as ``metastable'', readings. In the time domain this
question boils down to deciding about the precedence of events; it has been
extensively studied in context with flip-flops~\cite{KC87}, where a decision
needs to be made whether a transition on a data line occurred before or after a
given clock edge. Synchronizer circuits are used to reduce the risk
of metastable upsets to a suitable level, albeit at the cost of performance. In
case of the waiting synchronizer~\cite{Ginosar11} flip-flop stages are simply
cascaded to that end, which generally yields a dramatic improvement in reliability. 

In the value domain the most fundamental task is that of a discriminator which has to
decide whether a given input voltage is higher or lower than a reference. To
avoid oscillating behavior for inputs close to the reference, Schmitt-Trigger
(S/T) stages~\cite{FB94} are employed instead whose hysteresis behavior makes
them ignore irrelevant voltage fluctuations. However, it has been shown that for
certain input traces a S/T can become metastable as
well~\cite{Marino77, SMN16_async}. Following the lessons learned from synchronizers, one
may ask whether the cascading of S/Ts is again effective in reducing the risk of
metastable upsets. This is exactly the question we want to address in this
paper. 

To this end we will, in the next section, revisit the behavior of a single S/T
stage. Based on this knowledge we will, in Section~\ref{sec:questions}, break
down our key question into sub-questions whose answer may finally allow the
desired overall judgment. In Section~\ref{sec:model} we will briefly introduce
the metastability model for a single S/T stage, as derived by Marino in
~\cite{Marino77} and use it to shed light on our related sub-questions.  On the
foundation of this analysis we will elaborate predictions for the behavior of a
two-stage S/T cascade in Section~\ref{sec:claims} which we will validate by
extensive simulation experiments in Section~\ref{sec:validation}. Finally, we
will conclude the paper with Section~\ref{sec:conclusion}.

\section{Behavior of a single Schmitt-Trigger stage}
\label{sec:single_stage}

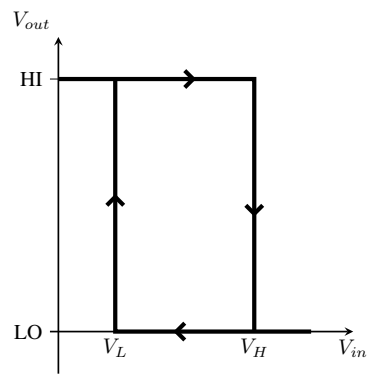
\begin{figure} [t]
  \centering
  \scalebox{0.8}{\begin{tikzpicture}[>=stealth, scale=0.7]

  \def\VDD{6}
  \def\XLIM{6}
  \def\VH{4.65}
  \def\VL{1.35}
  \def\VI{2.4}
  \def\VO{3.6}

  \def\aPosV{3}
  \def\aPosH{3}
  \def\aWidth{0.2}
  
  \draw[line width=0.8pt,->] (-0.2,0) -- (\XLIM+1,0) node[anchor=north] {$V_{in}$};
  \draw[line width=0.8pt,->] (0,-1) -- (0,\VDD+1) node[anchor=south east] {$V_{out}$};  

%  \draw[dashed] (-\lim,\valM) -- (\lim,\valM) node [anchor=south] {};
%  \draw[dashed] (-\lim,-\valM) -- (\lim,-\valM) node [anchor=north] {};

  \draw[black!100!white, line width=2pt] (0,\VDD) -- (\VH,\VDD) -- (\VH,0);
  \draw[black!100!white, line width=2pt] (\VDD,0) -- (\VL,0) -- (\VL,\VDD);

  \node[anchor=north] (VH) at (\VH,0) {$V_H$};
  \node[anchor=north] (VL) at (\VL,0) {$V_L$};

  \draw (0,\VDD) -- (-0.2,\VDD);
  \node[anchor=east] at (-0.2,\VDD) {HI};
  \node[anchor=east] at (-0.2,0) {LO};

%  \draw[line width=0.5pt, dashed] (\kHalf,-\valM) -- (-\kHalf,\valM);
 
  \draw[line width=2pt] (\aPosH,-\aWidth) -- (\aPosH-\aWidth,0) -- (\aPosH,\aWidth);
  \draw[line width=2pt] (\aPosH,\VDD+\aWidth) -- (\aPosH+\aWidth,\VDD) -- (\aPosH,\VDD-\aWidth);
 
  \draw[line width=2pt] (\VL-\aWidth,\aPosV) -- (\VL,\aPosV+\aWidth) -- (\VL+\aWidth,\aPosV);
  \draw[line width=2pt] (\VH+\aWidth,\aPosV) -- (\VH,\aPosV-\aWidth) -- (\VH-\aWidth,\aPosV);

\end{tikzpicture}}
\caption{S/T hysteresis}
\label{fig:hyst1}
\end{figure}

The key property of a S/T is its hysteresis behavior as shown in
Fig.~\ref{fig:hyst1}. Once an input voltage $V_{in}$ has crossed the upper
threshold $V_H$ from below, the S/T output flips to LO\footnote{We will consider
  an inverting S/T throughout the paper, as virtually all practical
  implementations are inverting.} and will not return to HI before the input
crosses the lower threshold $V_L$. This behavior provides the S/T its desired
robustness against fluctuations of $V_{in}$ which cause a comparator to
oscillate when occurring close to its (single!) threshold voltage. 

\begin{figure} [t]
%\tikzset{external/remake next}
%\tikzsetnextfilename{ST_circuit_ext}
% use circuitikz1
% http://www.texample.net/tikz/examples/tag/circuitikz/
  \centering
  \scalebox{1}{\includegraphics[width=0.99\linewidth]{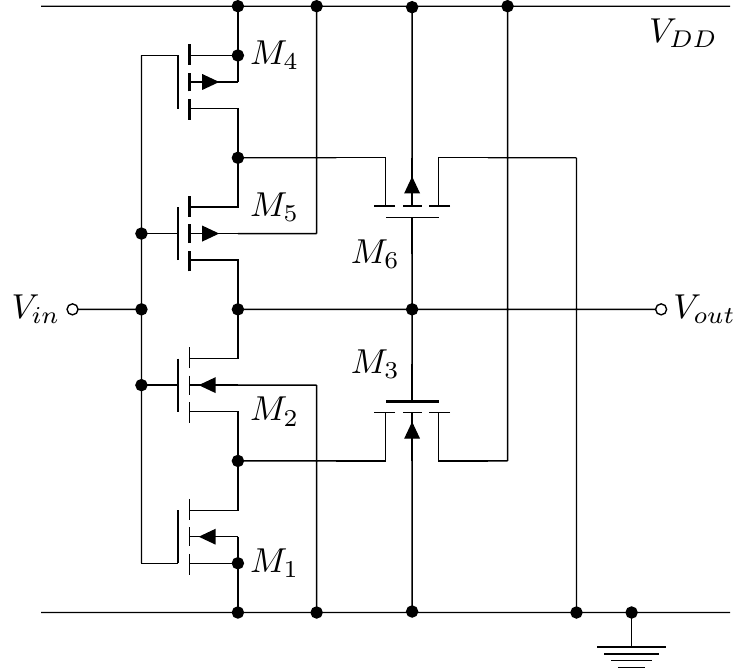}}
  \caption{Conventional CMOS \gls{st} implementation (from \cite{FB94})}
  \label{fig:CMOS_implementation}
\end{figure}

Clearly, the hysteresis behavior is tantamount to making the reference voltage
depend on the current output state, which, in turn, implies that there is some
kind of feedback of the output state to the
input. Fig.~\ref{fig:CMOS_implementation} shows a typical CMOS implementation of
a S/T \cite{FB94}. The feedback path via transistors $M_3$ and $M_6$ is clearly
visible.  It is exactly this (inevitable) existence of the positive feedback
path that makes the S/T prone to metastability. In case of an input trace that
first causes the S/T state to flip but then, before the feedback path has fully
stabilized, pulls the S/T back to the old state, undesired behavior can be
observed at the output. As will be elaborated in more detail in
Section~\ref{sec:model}, the S/T can exhibit the following behaviors at its
output:

\begin{enumerate}
\item[(B1)] For strictly monotonic input traces the S/T will show clean (steep) and
  well timed output transitions. This is the regular behavior.
\item[(B2)] A monotonic input trace that brings the S/T right to the tripping point
  and stays constant there can cause an indecision on whether to flip, which
  ultimately leads to a clean but (arbitrarily) late transition. 
\item[(B3)] Similarly, an input trace that goes slightly beyond the threshold but then
  back again and stays constant, can make the S/T assume a metastable state in
  which it outputs a constant voltage that is determined by the final constant
  value of $V_{in}$ and may well be somewhere in between a clean HI and a clean
  LO. 
\item[(B4)] An input pulse of marginal width, i.e. one that does not give the S/T
  enough time to stabilize its state, can create a glitch at the
  output. Depending on whether this pulse reaches full height or not, we
  distinguish between a glitch and a runt, respectively.  
\item[(B5)] In general, a non-monotonic input trace can, in principle, make the S/T
  output any desired voltage between clean HI and clean LO for any desired time
  -- in fact by appropriately controlling the input trace one can obtain any
  desired output voltage trace (within the S/T's static and dynamic voltage
  limits, of course). 
\end{enumerate}

It very much depends on the application which of these behaviors are actually
undesired, and as a consequence it is hard to quantify the improvement in signal
quality obtained by the S/T in general. In most cases marginal voltage levels
are problematic, as they could be interpreted differently by different
receivers. Unfortunately, in case (B3) one may still encounter marginal output
voltage levels that are not just transient, but, as these cases are hard to
trigger and hence can be considered rare, the S/T provides a very good
improvement in general. If output glitches are a concern, case (B4) becomes
problematic, and even in case (B2) a late transition may form a glitch in
conjunction with a regular subsequent one. Again, these cases need very specific
input traces and are hence rare.  Finally, due to case (B5), no type of output
behavior can be completely ruled out by using the S/T, which means there is only
a quantitative improvement, albeit a substantial one. However, as there are
infinite numbers of both, ``good'' and ``bad'' traces, it is not possible to
quantify the improvement; and we are not aware of any such approaches in the
literature.

\section{Questions to be answered}
\label{sec:questions}
From the above analysis we have concluded that a single S/T stage improves the
signal quality. So in principle, a subsequent S/T stage should obtain a similar
improvement, and thus overall the cascade has a higher gain than the single
stage. However, it remains to be investigated whether this is indeed the
case. In order to come to a conclusive answer, the following questions should be
addressed: 

\begin{enumerate}
\item[(Q1)] In which cases does the second S/T stage improve the behavior?
\item[(Q2)] Are there cases in which the behavior is getting worse? Are there new
  types of (likely) behavior? 
\item[(Q3)] Is the second stage equally likely to become metastable as the first one?
\item[(Q4)] Can metastability of the last stage be completely avoided, possibly by
  forming a longer cascade? 
\item[(Q5)] How are the static properties of the cascade determined (is it still a
  S/T, and if so, which hysteresis)? 
\item[(Q6)] How are the dynamic properties determined (regular delay, output slope, is
  there a performance penalty in using a cascade?) 
\item[(Q7)] Are there any rules for optimal dimensioning of the cascade (combination
  of fast and slow stage, different hystereses,...) 
\end{enumerate}

To conclusively answer these questions we will in the following elaborate a sufficiently detailed 
understanding of the behavior of a (2-stage) S/T cascade. As a first step towards this end we will 
revisit an existing model for the metastable behavior of a single S/T stage in the next section.

\section{Metastability model of a Schmitt-Trigger stage}
\label{sec:model}

\begin{figure}[t]
  \centering
  \scalebox{0.9}{\begin{tikzpicture}[scale=0.85, yscale=0.9]

  \draw
  (0,0) node[ocirc] (IN) {} to[short] ++(1,0) node[op amp, anchor=-] (opamp) {}
  (IN) node [anchor=east] {$V_{in}$}
%  (IN) to[R=$R_2$] ++(2.5,0) node[circ] (CONN1) {}
%  (CONN1) to[C,l_=$C_2$] ++(0,1.7) node[ground, yscale=-1] {}
%  (CONN1) to[short] ++(0.5,0) node[op amp, anchor=-] (opamp) {}
  (opamp.out) to[R=$R_0$] ++(2.7,0) node[circ] (CONN2) {}
  (CONN2) to[C,l_=$C_0$] ++(0,2.2) node[ground, yscale=-1] {}
  (CONN2) to[short] ++(1.4,0) node[ocirc] {} node[right] {$V_{out}$}
  (CONN2) to[R=$R_A$] ++(0,-1.9) node[circ] (CONN3) {}
  (CONN3) to[R=$R_B$] ++(0,-1.9) node[ocirc] {} node[below] {$V_R$}
  (opamp.+) to ++(-1,0) |- (CONN3)
%  (opamp.-) node [anchor=north east] {$V_{in}$}
%  (opamp.+) node [anchor=north east] {$V_1$}
%  (opamp.out) node [anchor=north west] {$V_3$}
  ;
  
\end{tikzpicture}}
  \caption{Dynamic model of the \gls{st} inspired by Marino \cite{Marino77}}
  \label{fig:Marino_Circuit}
\end{figure}
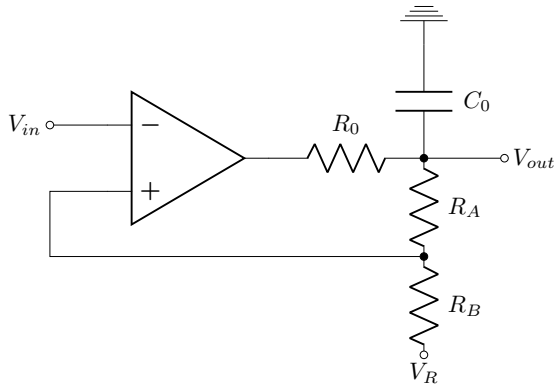

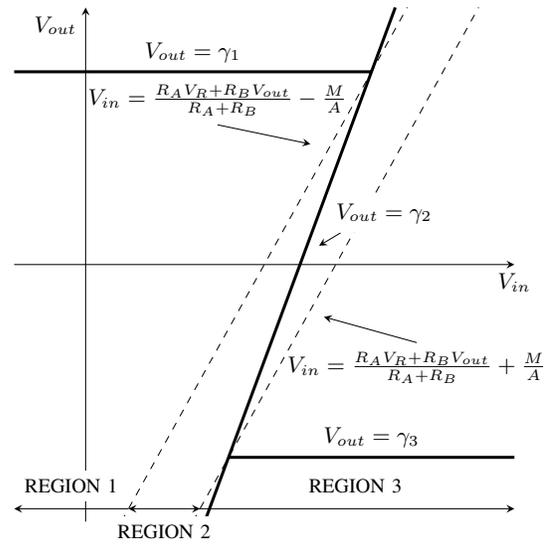
\begin{figure} [h]
  \centering
  \scalebox{0.95}{\begin{tikzpicture}[>=stealth, font=\small, yscale=0.9]

  \def\lim{4}
  \def\limNeg{-1}
  \def\valM{3}
  \def\VP{4}
  \def\VN{2}
  
  \draw[->] (\limNeg,0) -- (\lim+2,0) node[anchor=north] {$V_{in}$};
  \draw[->] (0,-\lim) -- (0,\lim) node[anchor=north east] {$V_{out}$};  

  \draw
  (0.5,-\lim)[dashed] -- node[pos=0.82, anchor=east] (L1)
  {$V_{in}=\frac{R_AV_R+R_BV_{out}}{R_A+R_B}-\frac{M}{A}$} (4.5,\lim) 
  (1.5,-\lim)[dashed] -- node[pos=0.3, anchor=west] (L3)
  {$V_{in}=\frac{R_AV_R+R_BV_{out}}{R_A+R_B}+\frac{M}{A}$}(5.5,\lim) 
  ;
  
  \draw[line width=1.2pt]
  (\limNeg,\valM) -- node[anchor=south] {$V_{out} = \gamma_1$} (\VP,\valM)
  (\VN,-\valM) -- node[anchor=south] {$V_{out} = \gamma_3$}(\lim+2,-\valM)
  (1.6666,-\lim) -- node[fill=white, anchor=west, pos=0.6, xshift=0.1cm] (L2) {$V_{out}=\gamma_2$}(4.3333,\lim)
  ;

  \draw[->] (L1.south) -- ++(1.3,-0.3);
  \draw[->] ([xshift=0.2cm]L2.south west) -- ++(-0.3,-0.2);
  \draw[->] (L3.north) -- ++(-1.3,0.4);
  
  \draw[<-] (\limNeg,-3.8) --
  node [yshift=0.1cm, pos=0.5,fill=white, anchor=south, font=\footnotesize] {REGION 1}(0.6,-3.8);
  \draw[<->] (0.6,-3.8) --
  node [yshift=-0.1cm, pos=0.5,fill=white, anchor=north, font=\footnotesize] {REGION 2} (1.6,-3.8);
  \draw[->] (1.55,-3.8) --
  node [yshift=0.1cm, pos=0.5,fill=white, anchor=south, font=\footnotesize] {REGION 3}(\lim+2,-3.8);
  
\end{tikzpicture}}
  \caption{Phase diagram for the \gls{st} from Marino \cite{Marino77}}
  \label{fig:Phase_Diagram}
\end{figure}

Marino has already proposed a dynamic model for the S/T that allows to
investigate its metastable behavior~\cite{Marino77}. This model is based on a
S/T implementation using an operational amplifier (OpAmp) as shown in
Fig.~\ref{fig:Marino_Circuit}. The OpAmp is assumed to be ideal, but with a
limited output voltage range of $\pm M$. An RC low pass at its output defines
its dynamic behavior, and a resistive voltage divider by $k \in \{0...1\}$ feeds
back part of the output voltage to the input (positive feedback). For brevity we
omit the derivation and solution of the associated differential equations here
(for details please refer to \cite{Marino77}) and just show the results:

As the saturation requires separate treatment, the solution comprises three
regions, as illustrated in Fig.~\ref{fig:Phase_Diagram}: upper and lower
saturation (Regions 1 and 3), as well as the ``linear region'' 2 between
them. The dashed lines represent the borders between the regions (the
corresponding equations are also given).

 The dynamic system represented by the S/T is then described by the following equations:
\begin{equation}
\label{eq:I_diff}
\text{Region~1:\hspace{1cm}}
\frac{dV_{out}}{dt} = V_{out}' = -\frac{1}{\tau_1} (V_{out} - \gamma_1)
\end{equation}
The resulting trajectory for $V_{out}$ is a decaying exponential function with time constant 
$\tau_1 \approx R_0 C_0$
that asymptotically approaches the truly stable rest point
$\gamma_1 \approx M$.

\begin{equation}
\label{eq:II_diff}
\text{Region~2:\hspace{1.4cm}}
\frac{dV_{out}}{dt} = V_{out}' = \frac{1}{\tau_2} (V_{out} - \gamma_2)
\end{equation}
Here we have a growing exponential function with time constant
$\tau_2 \approx \frac{R_0 C_0}{kA-1}$ that moves away from the metastable rest
point $\gamma_2 \approx \frac{V_{in}-(1-k)V_R}{k-\frac{1}{A}}$.  Note that we do
not have a single metastable point, as in case of a latch, but all points on
$\gamma_2$ are metastable points and the actual rest point depends on $V_{in}$
(see Fig.~\ref{fig:Phase_Diagram}).

\begin{equation}
\label{eq:III_diff}
\text{Region~3:\hspace{1cm}}
\frac{dV_{out}}{dt} = V_{out}' = -\frac{1}{\tau_3} (V_{out} - \gamma_3)
\end{equation}
Similar to Region~1 this yields a decaying exponential function with time constant 
$\tau_3=\tau_1 \approx R_0 C_0$
that asymptotically approaches the truly stable rest point
$\gamma_3 = -\gamma_1 \approx -M$.

At this point it is interesting to compare the S/T with a latch, as Veendrick
derived a similar model for latches in \cite{Vee80} to investigate their
metastability behavior. While both elements have a positive feedback, the key
difference is that in the latch the input becomes decoupled when the storage
loop is closed (i.e. when switching to hold mode). Mathematically this means
that the input becomes irrelevant, and just the homogeneous solution of the
differential equation applies. In the S/T this is not the case, and the input
continuously influences the behavior. This not only complicates the mathematical
treatment (and is probably the key reason why there are no quantitative
improvement values available for the S/T), it also results in the S/T having
more than one metastable point: depending on the input voltage the S/T can rest
in any metastable location along the $\gamma_2$ line and hence produce any
arbitrary metastable output voltage. Finally, this lack of decoupling between
stages also complicates the analysis of the S/T cascade. 

As a physical analogy one might consider the metastable latch as a stick that is
vertically placed on a firm table such that it balances on its tip before it
finally falls, while the S/T is the same vertical stick but balancing on a
finger that can still be moved left and right (inverted pendulum).

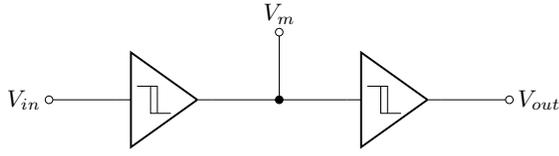
\begin{figure} [t]
  \centering
  \scalebox{0.9}{\begin{tikzpicture}

  \draw
  (0,0) node (VIN) {}
  (VIN) node[anchor=east] {$V_{in}$}
  (VIN) to[short,o-] ++(1,0) node[buffer,anchor=in] (ST1) {}
  (ST1.out) to[short,-*] ++ (1,0) node (CONN) {}
  (CONN) to [short,-o] ++ (0,1) node (VM) {}
  (VM) node[anchor=south] {$V_{m}$}
  (CONN) to [short] ++(1,0) node[buffer,anchor=in] (ST2) {}
  (ST2.out) to[short,-o] ++ (1,0) node (VOUT) {}
  (VOUT) node[anchor=west] {$V_{out}$}
  ;

  \draw (1.3,0.2) -| (1.6,-0.2);
  \draw (1.5,0.2) |- (1.8,-0.2);

  \draw (4.7,0.2) -| (5.0,-0.2);
  \draw (4.9,0.2) |- (5.2,-0.2);

\end{tikzpicture}}
\caption{Two stage \gls{st} cascade}
\label{fig:ST_cascade}
\end{figure}

\section{Claims about the behavior of a Schmitt-Trigger cascade}
\label{sec:claims}

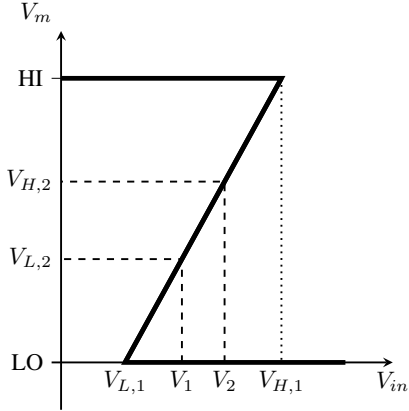
\begin{figure} [t]
  \centering
  \scalebox{0.9}{\begin{tikzpicture}[>=stealth, scale=0.7]

  \def\VDD{6}
  \def\VH{4.65}
  \def\VL{1.35}
  \def\VI{2.55}
  \def\VO{3.45}

  \def\VHT{4}
  \def\VLT{2}
  
  \def\aPosL{1}
  \def\aPosH{3}
  \def\aWidth{0.13}

  \draw[line width=0.8pt,->] (-0.2,0) -- (\VDD+1,0) node[anchor=north] {$V_{in}$};
  \draw[line width=0.8pt,->] (0,-1) -- (0,\VDD+1) node[anchor=south east] {$V_{m}$};  

  \draw[black!100!white, line width=2pt] (0,\VDD) -- (\VH,\VDD) -- 
  (\VL,0) -- (\VDD,0);

  \draw[black!100!white, line width=2pt, name path=char] (\VH,\VDD) -- (\VL,0);

%  \path[name path=vhline] (0,\VHT) node[anchor=east] {$V_{H2}$} -- +(\VDD,0);
%  \path[name path=vlline] (0,\VLT) node[anchor=east] {$V_{L2}$} -- +(\VDD,0);
  \node[anchor=north] at (\VL,0) {$V_{L,1}$};
  \node[anchor=north] at (\VH,0) {$V_{H,1}$};
  \draw[dotted, line width=0.8pt] (\VH,\VDD) -- (\VH,0);

  \path[name path=vhline] (\VI,0) node[anchor=north] {$V_{1}$} -- +(0,\VDD);
  \path[name path=vlline] (\VO,0) node[anchor=north] {$V_{2}$} -- +(0,\VDD);

  \draw (0,\VDD) -- (-0.2,\VDD);
  \node[anchor=east] at (-0.2,\VDD) {HI};
  \node[anchor=east] at (-0.2,0) {LO};

%  \draw[dashed, line width=0.8pt, name intersections={of=char and vhline, by=i}]
%    (0,\VHT) -- (i) -- (i |- 0,0) node[anchor=north] {$V_2$};

%  \draw[dashed, line width=0.8pt, name intersections={of=char and vlline, by=i}]
%    (0,\VLT) -- (i) -- (i |- 0,0) node[anchor=north] {$V_1$};

  \draw[dashed, line width=0.8pt, name intersections={of=char and vhline, by=i}]
    (\VI,0) -- (i) -- (i -| 0, 0) node[anchor=east] {$V_{L,2}$};

  \draw[dashed, line width=0.8pt, name intersections={of=char and vlline, by=i}]
    (\VO,0) -- (i) -- (i -| 0, 0) node[anchor=east] {$V_{H,2}$};

\end{tikzpicture}

%%% Local Variables:
%%% mode: latex
%%% TeX-master: "../paper"
%%% End:
}
\caption{Characteristic of a single S/T showing $V_1$ and $V_2$}
\label{fig:hyst_int}
\end{figure}

In the following we want to investigate a two stage \gls{st} cascade as shown in
Fig.~\ref{fig:ST_cascade}.  It was already argued in
Section~\ref{sec:single_stage} that a single S/T stage can essentially exhibit
any output behavior (case (B5)). This means that the first stage does not
qualitatively restrict the second one's input space, and, as a consequence,
stage 2 has unrestricted output behavior as well. So at this point we can
already answer question (Q4) about complete avoidance of metastability through a
S/T cascade: This is simply not possible.
 
In continuation of the physical analogy given in Section~\ref{sec:model} we can
view the second stage as a second vertical stick balancing on the upper tip of
the first one. This analogy nicely illustrates that it becomes much more
unlikely to see metastability in the second stage (i.e. actually find a balance
for the second stick) -- thus giving an intuitive answer to (Q3) -- , but it is
physically possible.

To get closer to a quantitative answer, let us analyze how the different output
behaviors of the first stage are handled by the second one. The internal signal
connecting the S/T is named $V_m$ (\emph{cf.}  Fig.~\ref{fig:ST_cascade}).

\subsection{Regular behavior}

Let us start with the regular behavior (B1) and assume a starting point with
$V_{in} < V_{L,1}$ (the case of $V_{in}>V_{H,1}$ is analogous). Note that in
$V_{L,i}$ and $V_{H,i}$ the index $i$ corresponds to the stage number.  Due to
the inverting behavior of each of our S/T stages, we have $V_m$ at HI and
$V_{out}$ at LO again.  As $V_{in}$ increases, $V_m$ and $V_{out}$ stay constant
until $V_{in}$ reaches $V_{H,1}$. Beyond that point $V_m$ will switch to
LO. With a strictly monotonic $V_{in}$ this transition of $V_m$ will be rapid.
Clearly, this transition of $V_m$ will also cause the second stage to switch,
namely when crossing its threshold $V_{L,2}$. Overall, we experience a clean
switching, with the threshold determined by that of the first stage, while the
second stage's threshold is irrelevant, as $V_m$ crosses the whole voltage range
anyway.  This answers question (Q5).

Generally, in this mode of operation we can expect the steepness of the
transitions to increase as the first \gls{st} tends to switch fast when its
threshold is reached, causing the second one to change even faster. The signal
is however delayed by the propagation delay of the second \gls{st}.

\begin{figure} [t]
  \centering
  \scalebox{0.9}{\begin{tikzpicture}[>=stealth, scale=0.7]

  \def\VDD{6}
  \def\XLIM{6}
  \def\VH{4.65}
  \def\VL{1.35}
  \def\VI{2.55}
  \def\VO{3.45}
  
  \def\aPosL{1.4}
  \def\aPosH{3.6}
  \def\aWidth{0.13}

  \draw[line width=0.8pt,->] (-0.2,0) -- (\XLIM+1,0) node[anchor=north] {$V_{in}$};
  \draw[line width=0.8pt,->] (0,-1) -- (0,\VDD+1) node[anchor=south east] {$V_{out}$};  

  \draw[black!100!white, line width=2pt] (0,0) -- (\VH,0) -- (\VI,0) --
  (\VO,\VDD) -- (\VL,\VDD) -- (\XLIM,\VDD);

  \draw[dashed, line width=0.6pt] (\VH,0) -- (\VH,\VDD);
  \draw[dashed, line width=0.6pt] (\VL,\VDD) -- (\VL,0);

  \draw[dotted, line width=0.6pt] (\VI,0) -- (\VI,\VDD);
  \draw[dotted, line width=0.6pt] (\VO,\VDD) -- (\VO,0);

  \node[anchor=north] at (\VH,0) {$V_{H,1}$};
  \node[anchor=north] at (\VL,0) {$V_{L,1}$};
  \node[anchor=north] at (\VI,0) {$V_1$};
  \node[anchor=north] at (\VO,0) {$V_2$};

  \draw (0,\VDD) -- (-0.2,\VDD);
  \node[anchor=east] at (-0.2,\VDD) {HI};
  \node[anchor=east] at (-0.2,0) {LO};

  \draw[line width=1pt] (\VL-\aWidth,\aPosL) -- (\VL,\aPosL-\aWidth) -- (\VL+\aWidth,\aPosL);
  \draw[line width=1pt] (\VH-\aWidth,\aPosH) -- (\VH,\aPosH+\aWidth) -- (\VH+\aWidth,\aPosH);
 
  \draw[line width=1pt] (\VO-\aWidth,\aPosL) -- (\VO,\aPosL-\aWidth) -- (\VO+\aWidth,\aPosL);
  \draw[line width=1pt] (\VI-\aWidth,\aPosH) -- (\VI,\aPosH+\aWidth) -- (\VI+\aWidth,\aPosH);

\end{tikzpicture}}
\caption{Theoretical hysteresis ($V_{in}-V_{out}$) of S/T cascade}
\label{fig:hyst}
\end{figure}
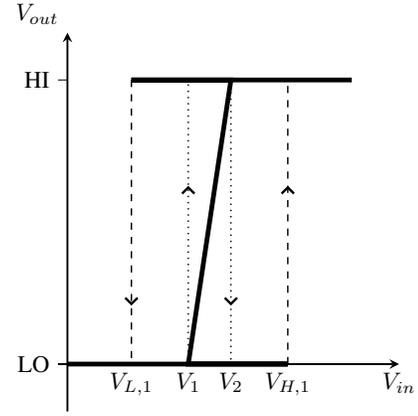

\subsection{Late transitions of stage 1}

According to case (B2) a ramp input stopping at a constant value near the
threshold will cause a late but clean transition at stage 1.  In that case the
second stage perceives a clean input which it simply conveys (adding its nominal
propagation delay). So late transitions are essentially not modified by the
second stage.

\subsection{Hysteresis curve of the cascade}

We know from Section~\ref{sec:single_stage} that for a single S/T to become
metastable its input voltage must be between its thresholds.  This means we can
only make the second stage metastable with a $V_m$ between $V_{L,2}$ and
$V_{H,2}$, which will (apart from a steep transition during switching as
described above) only occur when the first stage is metastable, and even then
only in a specific range, as shown in Fig.~\ref{fig:hyst_int}. In the following
we will use $V_1$ and $V_2$ to denote those values of $V_{in}$ that, if the
first S/T is metastable, will cause it to output $V_m = V_{L,2}$ and $V_{H,2}$
respectively. This consideration allows us to draw the hysteresis curve for the
overall behavior of the cascade shown in Fig.~\ref{fig:hyst}: Still the static
switching points are determined by the first stage, but the range of possible
metastable behavior is limited to the range of $V_1 \leq V_{in} \leq V_2$, with
$V_1$ and $V_2$ being determined by the slope of the $\gamma_2$ line of the
first stage and the thresholds of the second stage.

The same hysteresis is achieved when using two non-inverting \glsdesc{st}s,
while combining an inverting S/T with a non inverting one, mirrors the
hysteresis around the line $V_{out} = (HI + LO) / 2$, independent of the
ordering (assuming equal $V_{L,i}$ and $V_{H,i}$ values).

For the case that the hystereses of the single \gls{st}s are not equal the
ordering is important, as $V_H$ and $V_L$ of the cascaded system are determined
by the first \gls{st} alone and $V_1$ and $V_2$ by both of them. Let $V_{Ni}$ be
$(V_{H,i}+V_{L,i})/2$. If $V_{N1}= V_{N2} = VDD/2$ the order does not have an
influence on $V_1$ and $V_2$. In all other cases they might deviate, however the
difference between them, i.e. $V_2 - V_1$, is constant. This is very important
since $V_2 -V_1 < V_H - V_L$. Therefore it is possible, if the first \gls{st} is
held in metastability, to create a pulse train at the output with the reduced
hysteresis $V_1$ to $V_2$ (cycling the dotted lines in Fig.~\ref{fig:hyst}).

\subsection{Moving into metastability}

To move the second stage into metastability, stage 1 needs to be made metastable
first.  This can be attained by increasing $V_{in}$ from a value lower than
$V_{L,1}$ to $V_{H,1}$ and then decreasing it appropriately, while stage 1
starts to switch (behavior (B3)).  By staying close to the metastable restpoints
($\gamma_2$ in Fig.~\ref{fig:Phase_Diagram}) until $V_m$ reaches $V_{L,2}$
($V_{in} = V_1$) the second \gls{st} will start to switch and can be driven into
metastability in the same way as the first one.  While keeping both \gls{st}s in
metastability, which is either possible with very precise or very fast control
of $V_{in}$ (for a more detailed explanation see \cite{SMN16_async}),
the output values between LO and HI shown in Fig.~\ref{fig:hyst} are
reachable. 

\subsection{Emergence of glitches}

An interesting behavior can be observed when, with both stages in the metastable
state, $V_{in}$ is increased to a value between $V_2$ and $V_H$. This will bring
$V_m$ to above $V_H$ and hence cause the second stage to flip to LO. A further
(monotonic!) increase of $V_{in}$ beyond $V_H$ will then bring the first stage
to saturation and make $V_m$ transition to LO, which, in turn, causes the second
stage to flip back to HI. In this case we have observed a glitch at $V_{out}$
that was caused by a monotonic transition of $V_{in}$ (however, a non-monotonic
$V_{in}$ was initially required to bring both S/Ts into the metastable state in
the first place).  The root of this behavior lies in the positive slope of the
$\gamma_2$ line, which is somehow contradictory to the otherwise inverting
behavior of the S/T. This is also expressed by the direction of the arrows in
Fig.~\ref{fig:hyst}.

While this definitely represents a new type of (often undesired) behavior not
seen with a single stage -- thus answering (Q2) positively -- one should keep in
mind that it takes an \textit{extremely} precise control of $V_{in}$ to navigate
into this case. In the physical analogy this would equal the case of bringing
both sticks in vertical balance and then having them fall down to opposite
sides.

\subsection{Generalization of the analysis}

The above scenarios only represent some selected possible cases of output
behavior.  More generally, the shape of $V_{out}(t)$ is determined by (a) the
shape of the input voltage $V_{in}(t)$, (b) the state of stage 1 (given by
$V_m(t)$), and (c) the state of stage 2 (given by $V_{out}(t)$).  In this
3-dimensional space one can identify 9 characteristic sub-spaces (for $V_{in}$
remaining constant at its initial value). Five of these are listed in
Table~\ref{tab:stupid}, the other 4 are symmetric and not treated here for
brevity.  The subtables \tabNum{1}...\tabNum{5} apply for different values of
$V_{in}$.  Table \tabNum{1} shows the case where $V_{in}$ is significantly lower
than $V_{L,1}$, such that $V_m$ is forced to HI and consequently $V_{out}$ to LO
immediately, irrespective of their initial state.  The zeros denote the final
state of the cascade's output. The arrows show what event could be observed at
the output. 0 means the output was initialized to zero and remains there,
0\,\tikz[baseline=-0.7ex]{\path pic {downfast};} means the output makes a fast
transition to 0 from wherever it was initialized to. The
0\,\tikz[baseline=-0.7ex]{\path pic {miniglitch};} shows that there is (very
limited) glitch potential if the nominal propagation delays of the two S/Ts
significantly vary.

\begin{table} [h]
\centering
\caption{Possible output behaviors with static input}
\label{tab:stupid}
\begin{tikzpicture}[x=6mm, y=-4mm]

\def\mark{0.2}

\def\VDD{12}
\def\VH{9.3}
\def\VL{2.7}
\def\VI{5.1}
\def\VO{6.9}

\def\shift{-0.15}

\begin{scope}[line width = 1pt]
\draw[->] (0,0) -- (13,0) node [anchor=north] {$V_{in}$};
\draw (0,\mark) -- (0,-\mark) node [anchor=north, yshift=\shift cm] {$0$};
\draw (\VL,\mark) -- (\VL,-\mark) node [anchor=north, yshift=\shift cm] {$V_{L,1}$};
\draw (\VH,\mark) -- (\VH,-\mark) node [anchor=north, yshift=\shift cm] {$V_{H,1}$};
\draw (\VI,\mark) -- (\VI,-\mark) node [anchor=north, yshift=\shift cm] {$V_{1}$};
\draw (\VO,\mark) -- (\VO,-\mark) node [anchor=north, yshift=\shift cm] {$V_{2}$};
\end{scope}

\node[tablenum, anchor=south] at (0.3,-0.4) {1};
\node[tablenum, anchor=south] at (\VL-0.1,-0.4) {2};
\node[tablenum, anchor=south] at (3.9,-0.4) {3};
\node[tablenum, anchor=south] at (\VI-0.1,-0.4) {4};
\node[tablenum, anchor=south] at (6,-0.4) {5};

\begin{scope}[shift={(1,-3.5)}]
\path (-1,-1) node [tablenum] {1} (0,-1) node {0} (1,-1) node {1}
  (-1,0) node {1} (-1,1) node {0};
\path (0,0) node [cell] {} node {0};
\path (1,0) node [cell] {} node (t) [left=-3pt] {0} (t.east) pic {downfast};
\path (0,1) node [cell] {} node (t) [left=-2pt] {0} (t.east) pic {miniglitch};
\path (1,1) node [cell] {} node (t) [left=-3pt] {0} (t.east) pic {downfast};
\end{scope}
\begin{scope}[shift={(4.5,-3.5)}]
\path (-1,-1) node [tablenum] {2} (0,-1) node {0} (1,-1) node {1}
  (-1,0) node {1} (-1,1) node {0};
\path (0,0) node [cell] {} node {0};
\path (1,0) node [cell] {} node (t) [left=-3pt] {0} (t.east) pic {downfast};
\path (0,1) node [cell] {} node (t) [left=-2pt] {0} (t.east) pic {glitch};
\path (1,1) node [cell] {} node (t) [left=-3pt] {0} (t.east) pic {late};
\end{scope}
\begin{scope}[shift={(1,3)}]
\path (-1,-1) node [tablenum] {3} (0,-1) node {0}
  (1,-1) node {1} (-1,0) node {1} (-1,1) pic {metaup} (-1,2) pic {meta}
  (-1,3) pic {metadown};
\path (-1,4) node {0};
\path (0,0) node [cell] {} node {0};
\path (1,0) node [cell] {} node (t) [left=-3pt] {0} (t.east) pic {downfast};
\path (0,1) node [cell] {} node (t) [left=-2pt] {0} (t.east) pic {glitch};
\path (1,1) node [cell] {} node (t) [left=-3pt] {0} (t.east) pic {late};
\foreach \y in {2,3,4} {
  \path (0,\y) node [cell] {} node (t) [left=-3pt] {1} (t.east) pic {upfast};
  \path (1,\y) node [cell] {} node {1};
}
\end{scope}
\begin{scope}[shift={(4.5,3)}]
\path (-1,-1) node [tablenum] {4} (0,-1) node {0}
  (1,-1) node {1} (-1,0) node {1} (-1,1) pic {metaup} (-1,2) pic {meta}
  (-1,3) pic {metadown};
\path (-1,4) node {0};
\path (0,0) node [cell] {} node {0};
\path (1,0) node [cell] {} node (t) [left=-3pt] {0} (t.east) pic {downfast};
\path (0,1) node [cell] {} node (t) [left=-2pt] {0} (t.east) pic {double};
\path (1,1) node [cell] {} node (t) [left=-3pt] {0} (t.east) pic {late};
\foreach \y/\p in {2/uplate,3/uplate,4/upfast} {
  \path (0,\y) node [cell] {} node (t) [left=-2.5pt] {1} (t.east) pic {\p};
  \path (1,\y) node [cell] {} node {1};
}
\end{scope}
\begin{scope}[shift={(8,3)}]
\path (-1,-1) node [tablenum] {5} (0,-1) node {0} (1,-1) pic {metadown}
  (2,-1) pic {meta} (3,-1) pic {metaup}
  (-1,1) pic {metaup} (-1,2) pic {meta} (-1,3) pic {metadown};
\path (4,-1) node {1} (-1,0) node {1} (-1,4) node {0};
\path (4,-1) node {1};
\path (3,-1) pic {metaup};
\path (2,-1) pic {meta};
\path (1,-1) pic {metadown};
\path (0,-1) node {0};
\path (-1,0) node {1};
\path (-1,1) pic {metaup};
\path (-1,2) pic {meta};
\path (-1,3) pic {metadown};
\path (-1,4) node {0};
\foreach \x/\y/\sh/\val/\p in {
  0/0/4.5/0/nopic,4/0/3/0/downfast,
  0/1/4.5/0/nopic,1/1/2/0/metadown,2/1/2/0/metadown,3/1/2/0/doubledown,4/1/2/0/late,
  0/2/4.5/0/nopic,1/2/2/0/metadown,  3/2/2/1/metaup,4/2/5/1/nopic,
  0/3/2/1/uplate,1/3/2/1/doubleup,2/3/2/1/metaup,3/3/2/1/metaup,4/3/5/1/nopic,
  0/4/3/1/upfast,4/4/5/1/nopic} {
  \path (\x,\y) node [cell] {} node (t) [left=-\sh pt] {\val} (t.east) pic {\p};
}
\path (2,2) node [cell] {} pic {meta};
\end{scope}
\begin{scope}[shift={(11,-3.5)}]
\draw (0,0) grid [xstep=0.7, ystep=-0.7] (1.4,1.4);
\path (0.7,0) node [above] {$V_{out}$}
      (0,0.7) node [left] {$V_m$};
\end{scope}
\end{tikzpicture}

%%% Local Variables:
%%% mode: latex
%%% TeX-master: "../paper"
%%% End:
\end{table}

Table \tabNum{2} shows possible outputs when $V_{in}$ is so close to $V_{L,1}$
that it causes an abnormal switching delay of the first S/T (Region 2 in
Fig.~\ref{fig:Phase_Diagram}). Due to this delay, a late transition
(0\,\tikz[baseline=-0.7ex]{\path pic {late};}) or a glitch
(0\,\tikz[baseline=-0.7ex]{\path pic {glitch};}) can be observed at the output
when $V_m$ is initialized to $0$.  That would correspond to the case of glitch
emergence described above.

\begin{figure*} [t]
  \centering
  \includegraphics[trim=0 0 0 0,clip,width=1\linewidth]{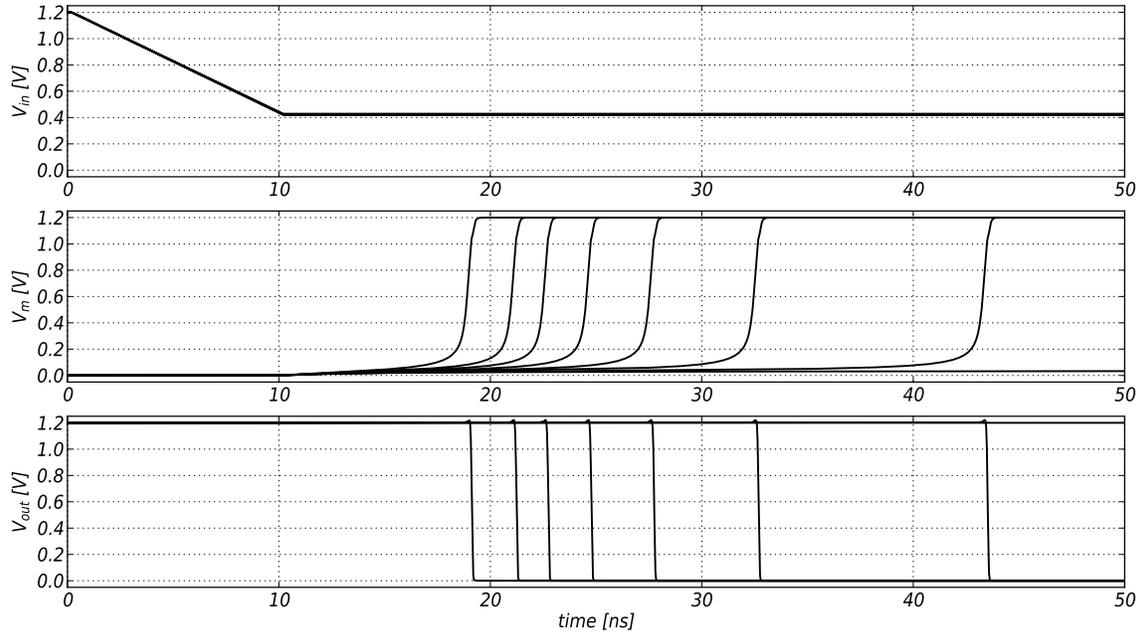}
\caption{Time trace of $V_{in}, V_m$ and $V_{out}$ for input slopes stopping at a
  constant value near $V_{L,1}$}
\label{fig:slope}
\end{figure*}

In table \tabNum{3}, $V_{in}$ is between $V_{L,1}$ and $V_1$ -- the first stage
can now also be initialized metastable resolving to 1
(\tikz[baseline=-0.7ex]{\path pic {metaup};}), to 0
(\tikz[baseline=-0.7ex]{\path pic {metadown};}) or not resolving at all
(\tikz[baseline=-0.7ex]{\path pic {meta};}). As can be seen, due to
$V_{in}<V_{L,1}$ the metastable voltage must be below $V_{L,2}$ for the next
stage, which can therefore not distinguish this type of metastability from a
clean LO.  When metastability resolves to HI, the effect on the second stage
will be identical to the late transitions in the previous paragraph.

Table \tabNum{4} shows the case when the metastable voltage output of the first stage
is very close to $V_{L,2}$. The difference to the previous table is that during
metastability of the first S/T, the second one can delay its output transition. The most
interesting case is 0\,\tikz[baseline=-0.7ex]{\path pic {double};}\,. It shows
that the relationship of the delays (metastability/late transition) makes a
qualitative difference at the output: If $V_m$ reaches HI first, the output
remains LO; if $V_{out}$ is faster, the output could be an arbitrarily delayed
glitch. 

Table \tabNum{5} presents the behaviors for input voltages of
$V_1<V_{in}<V_2$. Here both S/Ts can be initialized metastable, therefore the
table has many entries. Apart from late transitions,
0\,\tikz[baseline=-0.7ex]{\path pic {doubledown};} and
1\,\tikz[baseline=-0.7ex]{\path pic {doubleup};} again show the interesting
cases of race conditions. Depending on the order of metastability resolution,
the output can either go directly from an undefined voltage to HI or LO (first
S/T resolves before second one), or first go to the opposite logic state
followed by a full range transition (second S/T resolves before first one), thus
generating a glitch.  So we can conclude that in addition to the 3-dimensional
space described so far, the order of metastability resolution is important as
well, if both S/Ts are metastable.

As a general trend we can observe in the table that, while some of the
metastable cases of stage 1 are propagated, others are turned into other
behaviors like proper transitions, delayed transitions, or glitches. In no case
an intermediate voltage is generated from clean transitions.  So in applications
where the key purpose of using a S/T lies in protecting the subsequent logic
from intermediate voltages, the cascade does a decent job in reducing that
risk. However, if glitches and badly timed transitions are dangerous, then the
use of the cascade may be counter-productive. This may be considered an answer
to (Q1) ... (Q3).

\subsection{Pulse propagation}

According to case (B4) a single S/T stage may or may not propagate a glitch or
runt.  As the second stage may, of course, show the same behavior, glitches and
runts may propagate through the whole cascade. As the second stage may turn some
of these into stable transitions, however, the probability of propagation can be
expected to become lower in the cascade.

\section{Validation of the claims}
\label{sec:validation}

To validate our predictions about the behavior of the cascade from the previous
section, we performed HSPICE simulations of two S/Ts in series, each implemented
using the circuit shown in Figure~\ref{fig:CMOS_implementation} with transistor
parameters of an industrial 65\,nm process.

Fig.~\ref{fig:slope} shows the behavior of the cascaded \gls{st}s when a ramp
resulting in a constant value is applied to the input (case (B2)). As one can
see the first stage responds with late but clean transitions, i.e. ones that
cross the intermediate voltage range sufficiently fast. The second stage then
increases the steepness of the transitions even further. Please note that the
inputs causing the different traces only deviate by a very small amount of their
final, constant value. Therefore we can confirm, that $V_{H,1}$ respectively
$V_{L,1}$ has to be approached very accurately to observe late transitions at
the output. However, this is due to the properties of the first stage alone,
while the second stage does not yield any further improvement.

\begin{figure} [t]
  \centering
  \includegraphics[trim=0 0 0 2mm,clip,width=1\linewidth]{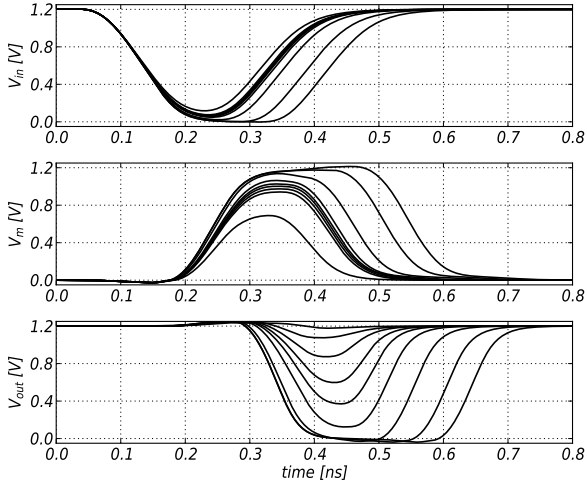}
\caption{Time trace of $V_{in}, V_m$ and $V_{out}$ for pulse inputs}
\label{fig:pulse}
\end{figure}

When short pulses are applied to the cascaded \gls{st}s (case (B4)), these are
preserved or suppressed depending on their width, as can be seen in
Fig.~\ref{fig:pulse}. A little bit misleading is the fact, that the pulses at
the output seem to be longer than those at the input, since the time they spend
below $V_L$ is longer than for those at the input. This can, however, be
explained by the increased steepness of the transitions. When the pulses are
compared by their crossing times of $VDD/2$, for example, one can see that they
stayed the same and are not stretched in time.

\begin{figure} [t]
  \centering
  \includegraphics[trim=0 0 0 0,clip,width=0.97\linewidth]{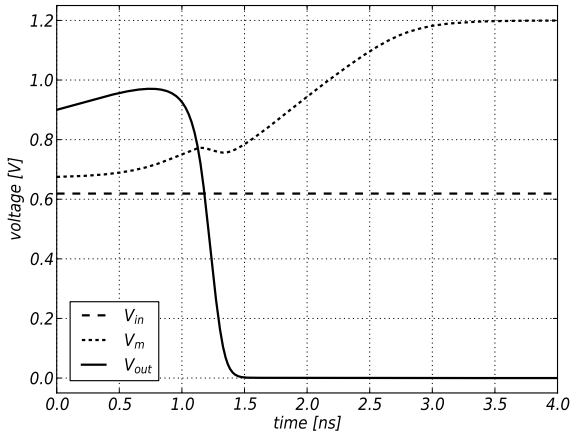}
\caption{First stage resolving earlier from metastability}
\label{fig:1stResolving}
\end{figure}

%\vspace{-1cm}

\begin{figure} [t]
  \centering
  \includegraphics[trim=0 0 0 0,clip,width=0.97\linewidth]{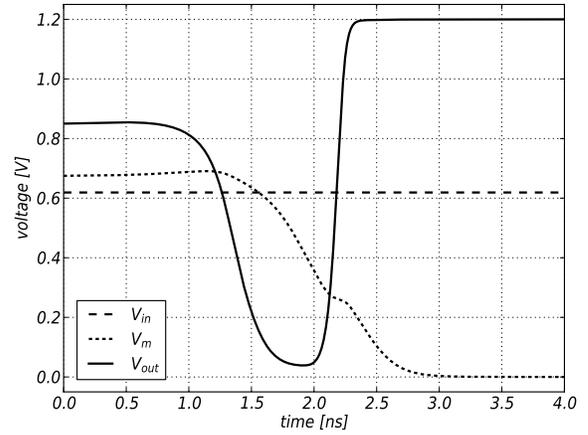}
\caption{Second stage resolving earlier from metastability}
\label{fig:2ndResolving}
\end{figure}

\begin{figure} [h!]
  \centering
  \includegraphics[trim=0 0 0 0,clip,width=0.97\linewidth]{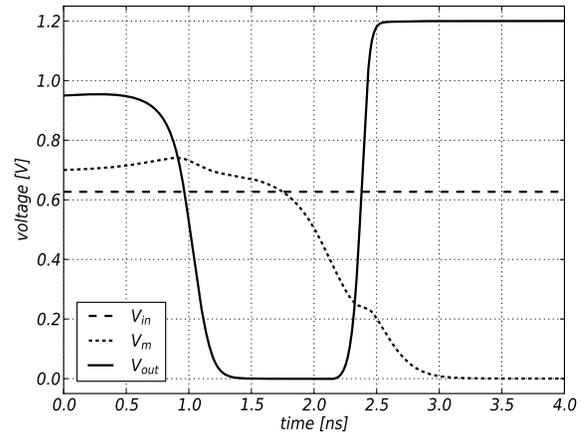}
\caption{Interactions between switching S/T stages}
\label{fig:influence}
\end{figure}

\begin{figure*} [t!]
  \includegraphics[width=\linewidth]{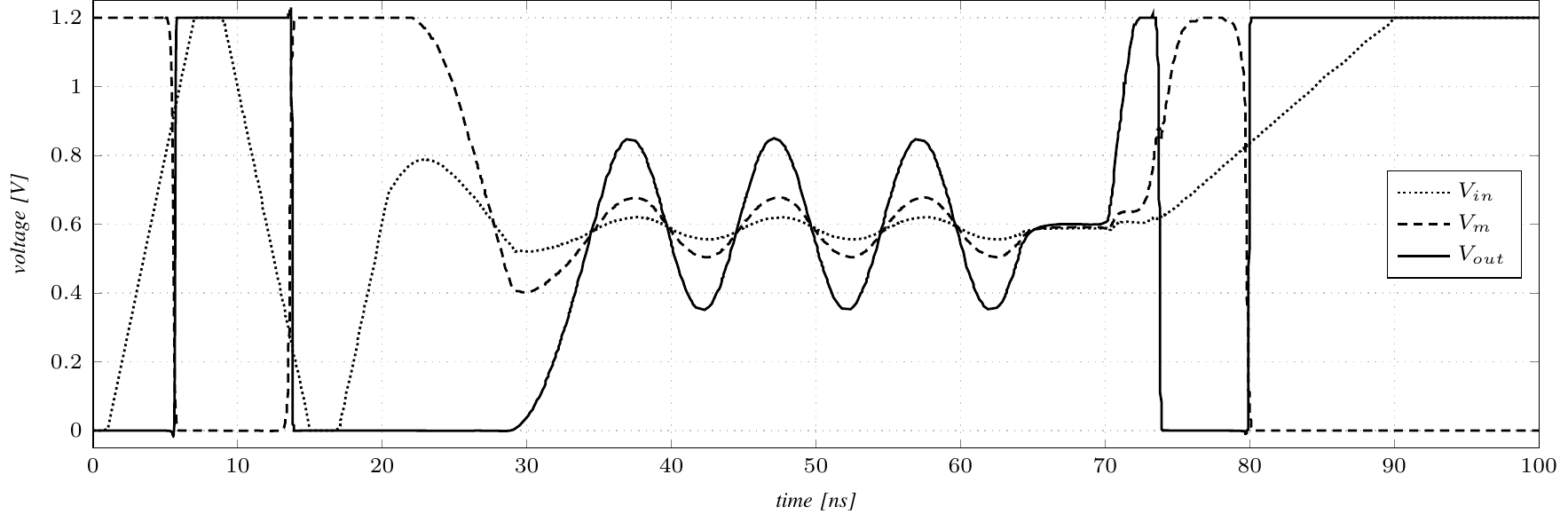}
\caption{``Arbitrary'' waveform created by operating both S/Ts in metastability}
\label{fig:arbitrarywf}
\end{figure*}

\begin{figure} [t]
  \includegraphics[width=\linewidth]{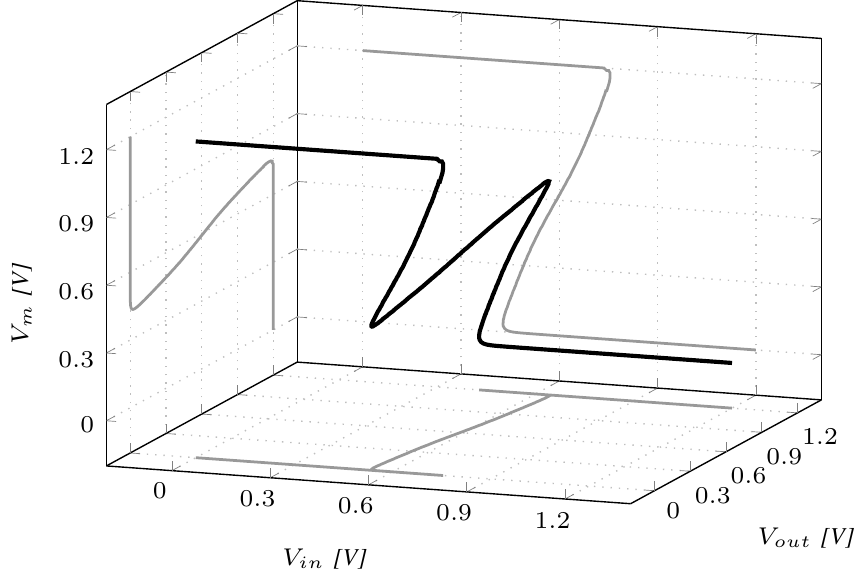}
  \caption{Characteristic of the cascade}
  \label{fig:char3D}
\end{figure}

Figures~\ref{fig:1stResolving} and \ref{fig:2ndResolving} show the cases when
both \gls{st}s are in their metastable region and the first respectively second
one resolves before the other. In the first case $V_{out}$ increases with $V_m$
until it reaches $V_H$, where $V_{out}$ drops rapidly. If, however, the second
\gls{st} resolves first and afterwards the first one to the same value, a glitch
at the output is introduced. This exactly matches our predictions from theory.

In the simulation we also observed a behavior that was not predicted, namely
that transitions of stage 2 influence the behavior of stage one.  This coupling
can be seen best in Fig.~\ref{fig:influence}. Initially the first \glsdesc{st}
resolves towards VDD until $V_{H,2}$ is reached, causing the second one to
switch. This, however, also introduces a change at the output of stage one
forcing it to drop.

Fig.~\ref{fig:arbitrarywf} shows how the S/T cascade can be driven into
metastability and even forced to output arbitrary waveforms (in this example a
sine wave). The beginning of the figure (first 15\,ns) shows regular transitions
(case (B1)). The hysteresis is clearly observable and it also becomes clear why
the cascade's thresholds/hysteresis are determined by those of the first S/T in
the series -- it applies its hysteresis to the input and outputs fast, full
range transitions. With such transitions at its input, the second S/T's
hysteresis only marginally increases the propagation delay through the cascade
but does not alter the overall hysteresis.

The next part of the figure (until 28\,ns) shows how the first S/T is driven
into metastability by carefully reverting $V_{in}$ when $V_m$ begins to
switch. The waveform that the metastable S/T outputs is such that the second S/T
becomes metastable in the same manner (from 28\,ns onward). With both S/Ts
metastable, the cascade is driven to output a sine wave (case (B5)) followed by
a constant output voltage (case (B3)). During the sine output the non-inverting
behavior of the single S/T stages and the amplification of $V_{in} to V_m$ and
then $V_{out}$ can be observed very clearly.  In the end (from 70\,ns) the
second S/T is resolving to VDD after which the first S/T also resolves to VDD,
forcing $V_{out}$ to GND. The input is continuously increased and eventually
crosses $V_{H,1}$, causing one last output transition -- again as predicted in
theory.

Finally, Fig.~\ref{fig:char3D} depicts the measured stable (truly stable plus
metastable) points in the ($V_{in}, V_m, V_{out}$)-space. The gray line at the
back is the projection of the 3D-curve to the plane $V_m$ over $V_{in}$ and thus
represents the hysteresis curve of the first stage.  The projection to the left
plane ($V_m$ over $V_{out}$) shows (a rotated version of) the hysteresis curve
of the second stage. The most interesting one is the projection of the curve to
the ground plane, i.e. $V_{out}$ over $V_{in}$. This represents the overall
hysteresis of the cascade. And indeed this curve exactly matches our prediction
from Fig.~\ref{fig:hyst}.

The 3D view also gives a better understanding of the reachability of the
metastable states with $V_{out}$ neither HI nor LO. These are only the states on
the line connecting $(V_{in}, V_m, V_{out}) = (0.64,0.74,1.2)$\,V and
$(0.53,0.43,0)$\,V. This line segment has two noteworthy properties: 1) Both its
projections to the first and second S/T characteristic coincide with metastable
states -- the output can only be held at an intermediate voltage when
\emph{both} S/Ts are metastable. 2) It does not start from any stable
state. This is not apparent in the 2D overall cascade characteristic. To reach
any point from that segment in a controlled manner, the first S/T has to be
metastable for a considerable time to produce the non-monotonic waveform
required to make the second S/T metastable. Intuitively, both arguments lead to
a substantially lower probability of the S/T cascade to output an intermediate
voltage compared to a single S/T.

\section{Conclusion}
\label{sec:conclusion}
We have addressed the question whether the cascading of Schmitt-Trigger stages
improves the metastable behavior. Our comprehensive theoretical and
simulation-based analysis showed that the risk of an intermediate output voltage
is indeed decreased, and generally the probability of metastable behavior is
significantly reduced. This is due to the fact that the first stage must
necessarily be metastable for the second stage to become metastable as
well. However, there are cases in which the interaction of the two states of the
two S/T stages causes extra transitions, and also some cases of intermediate
voltage are converted to transitions by the second stage. If the resulting,
potentially misplaced transitions are a problem, the usefulness of the cascade
should be carefully reconsidered in the given application context.

Another result is that metastable output behavior cannot be safely ruled out,
even with an arbitrarily long cascade. We have shown for the two-stage cascade
that an arbitrary output waveform can be generated, if only the input is
controlled precisely enough.


\end{document}